\documentclass[namedreferences]{solarphysics}

\usepackage[hyperref,optionalrh]{spr-sola-addons-arxiv} 
\usepackage{graphicx}        
\usepackage{color}           
\usepackage{multicol}
\usepackage[flushleft]{threeparttable}
\usepackage[utf8]{inputenc}
\usepackage{multirow}
\usepackage{booktabs}
\usepackage{siunitx}
\usepackage{mathptmx}
\usepackage{balance}
\usepackage{todonotes}
\usepackage{enumerate}
\usepackage{hyperref}
\usepackage{framed}

\chardef\us=`\_

\colorlet{shadecolor}{yellow}
\begin{document}

\begin{article}
\begin{opening}

\title{Solar Flares Forecasting Using Time Series and Extreme Gradient Boosting Ensembles}

\author[addressref={aff1,aff2},corref,email={tiago.cinto@pos.ft.unicamp.br}]{\inits{T.}\fnm{T.}~\lnm{Cinto}}
\author[addressref=aff1,email={gradvohl@ft.unicamp.br}]{\inits{A. L. S.}\fnm{A. L. S.}~\lnm{Gradvohl}}
\author[addressref=aff1,email={guilherme@ft.unicamp.br}]{\inits{G. P.}\fnm{G. P.}~\lnm{Coelho}}
\author[addressref=aff1,email={aeasilva@ft.unicamp.br}]{\inits{A. E. A.}\fnm{A. E. A.}~\lnm{da Silva}}


\address[id=aff1]{School of Technology (FT), University of Campinas (UNICAMP), Limeira, SP, Brazil}
\address[id=aff2]{Federal Institute of Education, Science and Technology of Rio Grande do Sul (IFRS), Campus Feliz, RS, Brazil}

\runningauthor{Cinto et al.}
\runningtitle{Solar Flares Forecasting Using Time Series and Extreme Gradient Boosting Ensembles}

\begin{abstract} 
Space weather events may cause damage to several fields, including aviation, satellites, oil and gas industries, and electrical systems, leading to economic and commercial losses. Solar flares are one of the most significant events, and refer to sudden radiation releases that can affect the Earth's atmosphere within a few hours or minutes. Therefore, it is worth designing high-performance systems for forecasting such events. Although in the literature there are many approaches for flare forecasting, there is still a lack of consensus concerning the techniques used for designing these systems. Seeking to establish some standardization while designing flare predictors, in this study we propose a novel methodology for designing such predictors, further validated with extreme gradient boosting tree classifiers and time series. This methodology relies on the following well-defined machine learning based pipeline: (i) univariate feature selection provided with the F-score; (ii) randomized hyper-parameter search and optimization; (iii) imbalanced data treatment through cost function analysis of classifiers; (iv) adjustment of cut-off point of classifiers seeking to find the optimal relationship between hit rate and precision; and (v) evaluation under operational settings. To verify our methodology effectiveness, we designed and evaluated three proof-of-concept models for forecasting $\geq C$ class flares up to 72 hours ahead. Compared to baseline models, those models were able to significantly increase their scores of true skill statistics (TSS) under operational forecasting scenarios by 0.37 (predicting flares in the next 24 hours), 0.13 (predicting flares within 24-48 hours), and 0.36 (predicting flares within 48-72 hours). Besides increasing TSS, the methodology also led to significant increases in the area under the ROC curve, corroborating that we improved the positive and negative recalls of classifiers while decreasing the number of false alarms.
\end{abstract}
\keywords{Flares, Forecasting; Flares, Models; Active Regions, Magnetic Fields; Active Regions, Structure.}
\end{opening}

\section{Introduction}
\label{sec:introduction} 

The Sun's atmosphere is very active and features several events that directly affect the solar system bodies. These events can affect the solar wind, the Earth's atmosphere, and also the near-Earth space, which can lead to strong effects on oil and gas industries, satellites, aviation, and electrical systems \citep{NRC:2009}.

Mostly related to active regions (ARs) \citep{Canfield:2001}, solar flares are sudden releases of x-rays in the 1-8 \AA{ngstr{\"o}m} (\si{\angstrom}) wavelength, measured in watts per square meter (\si{\watt\per{\square\meter}}) \citep{Messerotti:2009}, recorded by the Geostationary Operational Environmental Satellite (GOES) x-ray instrument. To represent the magnitude of a flare event, those releases are classified through a labeled scale ranging between A-, B-, C-, M-, and X-class. 

Each flare class has its x-ray peak flux ten times higher than its predecessor. Besides, each class varies according to a linear scale ranging from 1 to 9, which represents the flare intensity. Thus, to report flares, we must multiply their class peak values by their corresponding intensity factors. 

Flare releases can affect the Earth's atmosphere in a few hours or minutes, and thus are considered one of the most significant events of the Sun. Because of the several reported effects of solar flares, it is imperative to employ efforts for forecasting them. 

%
%
%
%

\subsection{Solar Flares Forecasting Efforts}

Many studies use data on the photospheric magnetic field to evaluate the relationship between ARs and solar activity \citep{McAteer:2010}. Others focus on the photospheric features \citep{McIntosh:1990} and magnetic topologies \citep{Hale:1919} of ARs, which are known to be associated with solar flares productivity. Regardless of the guiding characteristic, researchers designed many methods to forecast solar flares, including the following most notable examples: expert systems \citep{McIntosh:1990, Crown:2012, Murray:2017, Devos:2014}; Poisson statistics \citep{Bloomfield:2012}; random forest \citep{Domijan:2019}; linear discriminant analysis \citep{Leka:2018}; support-vector machines \citep{Yang:2013, Muranushi:2015}; relevance-vector machine \citep{Al-Ghraibah:2015a}; multiple linear regression \citep{Shin:2016}; neural networks \citep{Colak:2009, Ahmed:2013, Hada-Muranushi:2016, Nishizuka:2018, Huang:2018}; radial basis function networks \citep{Qahwaji:2007}; and regression models \citep{Anastasiadis:2017,Song:2009}. Besides, \cite{Leka:2019a} and \cite{Leka:2019b} present other examples. 

Despite the benefits and improvements in the performance of the aforementioned systems, most of them share a common aspect: the use of statistical or machine learning techniques to design their prediction models. Machine learning is a recurrent domain of computer science whose techniques can be used to learn from historical data and forecast future observations (supervised learning task) \citep{Han:2006}. To train such algorithms, we must provide them with training samples and their corresponding labels representing the existence of some specific event. 

Besides, to design effective forecasting systems, we must pay careful attention to the used methodology. Theoretically, powerful learning algorithms would become useless if they did not follow a strict and well-structured methodology. However, usually, methodologies are tailor-made and designed for specific purposes, i.e., they are created and adjusted for the specific scenarios of each study, thus restricting their use for automated space weather forecast. 

\subsection{Automated Space Weather Forecast}
By automated space weather forecast, we mean platforms that rely on generic -- yet flexible -- methodologies to easily design and evaluate new forecasting models when necessary. This need for new models can arise with the availability of new data, prediction of several events other than solar flares, or even be linked to a production environment for real-time forecasting. Besides flexible, these methodologies must also provide some standardization while designing flare predictors, i.e., their processes must effectively adhere to most of the available learning algorithms and flare concerns. 

To date, there are a few studies whose authors focused on proposing some standardization while designing flare forecasting systems despite the numerous predictors already published. To name some, we can mention the studies by \cite{Muranushi:2015} and \cite{Leka:2018}.

\cite{Muranushi:2015} proposed the Universal Forecast Constructor by Optimized Regression of Inputs (UFCORIN). Using linear regression, their system comprehended a generic time series predictor, which could be set to forecast any time series feature from an arbitrary set of input data. 

UFCORIN provide users with the flexibility to easily change the input and output target features when new data became available or new models were needed. Despite its flexibility, the approach had some disadvantages, such as it was restricted to linear regression predictors.

On the other hand, \cite{Leka:2018} proposed the Discriminant Analysis Flare Forecasting System (DAFFS). DAFFS comprehended an automated solar flare prediction system for analyzing the Sun's magnetic fields, seeking evidence of stored energy and magnetic complexity associated with flare productivity. 

By using near-real-time vector magnetic data along with several reports from GOES, the DAFFS methodology trained new models on-demand to support real-time forecasting. This tool was executed twice a day to make predictions for several validity periods and event magnitudes (i.e., prediction of $\geq C1.0$, $\geq M1.0$, $\geq X1.0$ flares in the next 24 and 24-48 hours).

When designing new models, DAFFS always used as much data as possible; noteworthily, it considered the period between 2012 and the most recent month at the time of its execution. However, DAFFS did not only assist in the training of new models, but also provided several customization options for them, which could involve: custom flare magnitude threshold definition, precision vs. recall trade-off adjustment, type of prediction adjustment (full-disk or AR-by-AR), among others. 

\subsection{Aims and Scope}

The skill in customizing prediction models is paramount in automated space weather forecast. Not only customization, but also well-structured and robust training methodologies are required in such platforms. In this article we propose a pipeline of techniques to design, train, and evaluate flare forecasting systems under operational settings with flexibility and performance.

We define systems evaluated under operational settings as models that forecast truly unseen data, i.e., data not previously used during their design. In this sense, we evaluate models with novel data, thus also calling models designed under our methodology as ``operationally-evaluated''. 

This pipeline comprehends a methodology with some well-defined and connected processes, namely: automated feature selection, fine-tuning of hyper-parameters of \linebreak models, analysis of the cost function of classifiers, and adjustment of the precision vs. recall cut-off point of classifiers. This methodology will support a novel automated space weather forecast platform in the future. In addition, we shall present a case study to verify the effectiveness of this methodology.

We divided this paper into six sections. Besides the introduction of Section~\ref{sec:introduction}, we describe the dataset prepared to evaluate our methodology as well as our flare catalog and features in Section~\ref{sec:dataset}. In Section~\ref{sec:methodology}, we explain the proposed methodology, providing details of each performed step and employed techniques. In Section~\ref{sec:results}, we underlie our results and show how the methodology has improved the performance of the forecasting systems. In Section~\ref{sec:literature-analysis}, we compare our designed models and methodology with those addressed in the literature to verify the effectiveness of our approach. Finally, in Section 6, we underlie the conclusions of our research.

\section{Dataset}
\label{sec:dataset} 

This section shall present the data assembled to evaluate the herein proposed methodology. The Space Weather Prediction Center (SWPC) is a service center linked to the National Oceanic and Atmospheric Administration (NOAA)\footnote{http://www.swpc.noaa.gov}, located in Boulder, Colorado - USA. SWPC aims at climate prediction, and thus continually provides real-time monitoring of solar events that affect navigation, telecommunications, and satellites. As the official source for space weather alerts in the United States, data collected by NOAA/SWPC are freely available for research and study purposes\footnote{ftp://ftp.swpc.noaa.gov/pub/warehouse/}.

NOAA/SWPC keeps monitoring data into several distinct repositories, for instance: daily solar data (DSD), which comprises the Sun's behaviors and daily aggregated records, and sunspot region summary (SRS), which further details active regions re\-cord\-ed in DSD, i.e., it does provide their magnetic types, locations in the solar disk, corresponding areas etc. We downloaded and integrated data from those two NOAA/SWPC repositories at first. Since both DSD and SRS comprehend daily aggregated data and NOAA compile them once a day, we used their compilation dates (year/month/day) for integration purposes.

We considered the date period between January 01, 1997 and January 15, 2017. Overall, we assembled 7,320 records comprehending data from several distinct features\footnote{Data prepared in this section are available from https://doi.org/10.5281/zenodo.3517025}:

\begin{itemize}
\item \emph{X class flares}: number of daily observed X class flares (used for assembling the target variable) \citep{Noaa:2011}.
\item \emph{M class flares}: number of daily observed M class flares (used for assembling the target variable) \citep{Noaa:2011}.
\item \emph{C class flares}: number of daily observed C class flares (used for assembling the target variable) \citep{Noaa:2011}.
\item \emph{Sunspot area}: the daily sum of all sunspot areas \citep{Noaa:2011}.
\item \emph{Sunspot number}: the daily aggregated number of sunspots \citep{Noaa:2011}.
\item \emph{X-ray background flux}: the daily average background x-ray flux \citep{Noaa:2011}.
\item \emph{Radio flux}: the daily solar radio flux reported by the Dominion Radio Astrophysical Observatory at Penticton, Canada \citep{Noaa:2011}.
\item \emph{Daily weighted mean flare rate (WMFR) of magnetic classes}: Mt. Wilson classes (or magnetic ones) describe sunspots according to their magnetic complexities \citep{Hale:1919}. The WMFR, in turn, calculates the weighted mean rate of each Mt. Wilson class regarding its occurrence along with C-, M-, and X-class flares, and is defined as in Equation~\ref{eq:wmfr} \citep{Shin:2016}:
\begin{equation}
\label{eq:wmfr}
wmfr_{daily} = \sum_{i=1}^{y} \frac{n_{c}+10n_{m}+100n_{x}}{n}
\end{equation}
\noindent where $n_{c}$ is the number of times each Mt. Wilson class happened along with C-class flares; $n_{m}$ is the number of times each one was linked to M-class flares; $n_{x}$ is the number of times each one was linked to X-class flares; $n$ is the number of times each one was recorded regardless of being linked to flares; and the constants 10 and 100 are the weights for flare importance. Since our data refer to daily aggregated solar behaviors, several sunspots with different WMFRs may be recorded in one day. Thus, we consider the daily WMFR a summation of all calculated WMFRs on a given day ($y$ is the number of sunposts and $i$ represents the WMFR linked to each sunspot). We only calculated the WMFR of the most frequent magnetic classes based on \cite{Jaeggli:2016}'s research, in which they analyzed the years between 1992 and 2015 (most of the period we are considering), and highlighted the most frequent cases. The full list of considered magnetic classes included: \textit{alpha, beta, gamma, beta-gamma, gamma-delta, beta-delta}, and \textit{beta-gamma-delta}.
\item \emph{Daily WMFRs of the Zpc components of McIntosh classes}: McIntosh classes describe sunspots according to three components, namely \textit{Zpc} \citep{McIntosh:1990}. Some authors argue that these components alone can be used as proxies to represent the emergence or decay of magnetic fluxes inside sunspot groups \citep{Eren:2017,Kilcik:2018,McCloskey:2016}. Some cases of flux emergence directly lead to higher flare amounts, thus representing correlations between individual components and observed flares. The daily WMFR calculation for the components of McIntosh classes follows what we defined in Equation~\ref{eq:wmfr} for magnetic classes. However, it does consider each \textit{Zpc} component independently. Thus, we calculated three daily WMFR values for each sample of our data. 
\end{itemize}

\subsection{Data Pre-processing}
\label{sec:data-pre-processing}

We employed two distinct techniques for data preparation, namely missing data imputation and data standardization \citep{Han:2006}. Regarding missing data, examples of this issue include: dataset issues (i.e., DSD records without their corresponding SRS data), features with absent values (i.e., sunspot area and number), and also features with noisy data (i.e., x-ray background flux with zeroed measures).

We input missing data through the k-Nearest Neighbors algorithm ($k$-NN) \citep{Han:2006}, i.e., a technique with which we input data considering the similarity between data records. This similarity is defined in terms of a distance metric such as the ordinary Euclidean score. The Euclidean distance between two records $X_{1}=(x_{11}, x_{12}, ..., x_{1n})$ and $X_{2}=(x_{21}, x_{22}, ..., x_{2n})$ is defined as in Equation~\ref{eq:euclidean-distance} \citep{Zaki:2013}:

\begin{equation}
\label{eq:euclidean-distance}
dist(X_{1},X_{2})=\sqrt{\sum_{i=1}^{n}(x_{1i}-x_{2i})^2}
\end{equation}

\noindent where $n$ is the number of features; $x_{1i}$ is the first record \textit{i-iest} feature; and $x_{2i}$ is the second record \textit{i-iest} feature. To input a missing feature value, the $k$-NN uses the most predominant value among the $k$-neighbors that are the closest to the record with missing data.

Besides missing data, the dataset also had dissimilar data ranges among features. Whereas the sunspot number and area, and radio flux respectively ranged between \big[0; 401\big], \big[0; 5690\big], and \mbox{\big[65; 298\big]}, the background flux calculation of x-ray was ranging in a shorter interval, \mbox{\big[$10^{-7}$; $2\times10^{-5}$\big]}. This issue is known to highly affect the performance of machine learning predictors, in such a way we also considered the z-score normalization for features \citep{Han:2006}.

Also known as the standard score, the z-score is an algorithm that reduces ranges of data to $\mu=0$  and $\sigma=1$. The use of the z-score instead of an ordinary min-max normalization is highly recommended since it positively affects the classifiers predictive performance \citep{Nishizuka:2017}. In Equation~\ref{eq:z-score} we show how to calculate the z-score for each feature.

\begin{equation}
\label{eq:z-score}
z = \frac{x - \mu}{\sigma}
\end{equation}
\newline
\noindent where $\mu$ is the feature mean; $\sigma$ is the feature standard deviation; $x$ is an arbitrary feature value; and \emph{z} is the standardized value related to $x$.

\subsection{Sliding time window}
Learning systems that cope with the evolution of data over a time period and forecast events in a supervised and sequential way are defined short-term predictors~\citep{Yu:2009}. To provide proper data input for them, we should design data under the sliding time window schema, whose aim is to represent the evolution of data $n$ days before the existence of some event. Many authors argue in favor of using this time window combined with flares forecasting, noticeably \cite{Yu:2009} and \cite{Huang:2010}. 

With the sliding time window, solar data are always observed at the $t$ instant (an arbitrary day) and some days before $t$, i.e., $[t - \Delta t]$. The interval between $t$ and $[t - \Delta t]$ defines the sliding time window.

A key issue when designing data considering a sliding time window is how to choose a reasonable window length. Input data may not be enough when extremely short windows are designed. On the other hand, long windows may lead to useless data imputation.

In an attempt to deal with enough data, we defined our window length based on the lifetime of ARs. According to \cite{Canfield:2001}, active regions are born when magnetic flux strands become visible into the atmosphere from the solar interior. Usually, flux emergence continues for as long as 5 to 7 days, which refers to the period with which ARs grow to larger sizes and then flux quickly stops emerging until ARs vanish. Therefore, we designed our data stream regarding 5 days of evolutionary data, i.e., we aimed to cover ARs during their entire life cycles.


\subsection{Event Definition and Forecasting Horizons}

We use the sliding time window mentioned earlier to predict events that take place some periods ahead of the $t$ instant, i.e., forecasting horizons. We designed our event definition (target feature) as the occurrence of at least one C-, M-, or X-class flare within the next 24, 24-48, and 48-72 hours ahead of the $t$ instant. We defined the positive events according to NOAA/SWPC's events count.

Regarding the class ratio distribution, there were about 58\% of data referring to the positive class (i.e., 4,225 samples were linked to at least one C-class flare or higher). In contrast, the rest represented the absence of events (42\%, 3,095 samples).

We based our approach for predicting flares on The Met Office Space Weather Operations Centre (MOSWOC) in the United Kingdom \citep{Murray:2017} and also on NOAA/SWPC. Using a hybrid-forecasting technique, MOSWOC/SWPC make full-disk forecasts based on Poisson statistics, which are then adjusted by human experts. However, MOSWOC predicts specific classes of flares in the next 24, 24-48, and 48-72  hours instead of flares above a threshold as in this study. On the other hand, in NOAA/SWPC, the forecasting horizons are also designed as in MOSWOC, but they make forecasts for events higher than or equal to some magnitude thresholds, namely $\geq C$ and $\geq M$~\citep{Crown:2012}. 

Forecasting events within certain periods ahead of the $t$ instant is more realistic than forecasting them exactly after $t$~\citep{Jonas:2018}. In addition, since our dataset provides daily aggregated flare data, our models make full-disk flare forecasts regardless of which active region produces the event.

The reason for assembling the dataset as such is merely related to our case study. We do not restrict our proposed methodology to this type of data, and one can use it along with other features with only a few or no adjustments.

\section{Methodology}
\label{sec:methodology} 
In this section we describe the proposed methodology. Besides explaining its inner processes, we also detail some supporting concepts, such as the involved mathematics.

As a case study to assess the methodology effectiveness, we shall use extreme gradient boosting tree models (XGBoost) \citep{Chen:2016}. Such models belong to the ensemble class of machine learning algorithms. Ensembles are meta-algorithms that combine the output of several individual classifiers following certain rationale to optimize the predictive performance~\citep{Han:2006}.

Extreme gradient boosting tree models rely on the boosting strategy. Thus, this algorithm gradually and sequentially trains several individuals of the same type -- decision trees --, always attempting to minimize their loss functions \citep{Chen:2016}. Besides boosting, XGBoost also relies on the bagging strategy, according to which each individual is trained under a different subsample drew from the training set, thus reducing variance and helping to avoid over-fitting \citep{Hastie:2009}.

The reason for choosing the extreme gradient boosting tree algorithm is merely related to our case study. The proposed methodology is not restricted to such class of models in any way, and thus could be used along with other learning algorithms with only a few or even no adjustments.

\subsection{Data Splitting}
\label{sec:data-splitting}

\begin{figure*}[!htb]
\centering
\includegraphics[width=1.0\textwidth]{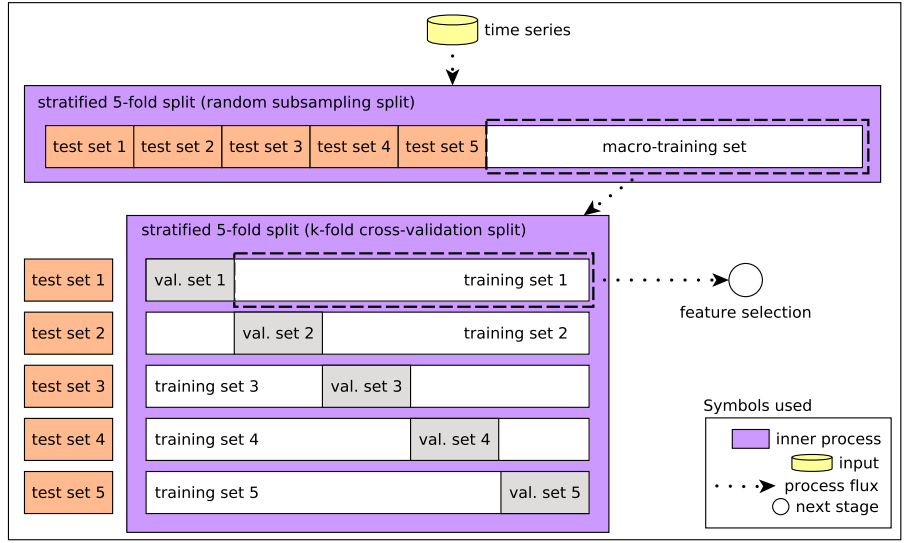}
\caption{Data splitting schema.}
\label{fig:data-splitting}
\end{figure*}

The methodology workflow starts with a dataset splitting into macro-training data and five test segments. Then, we re-split macro-training data into validation and training sets (Figure~\ref{fig:data-splitting}). Henceforth, let us define those concepts as follows:

\begin{itemize}
    \item \textit{Test data}: unseen data samples used to evaluate the model generalization error under operational settings;
    \item \textit{Validation data}: data samples similar to the test ones, used to verify whether the output model is ready to be operationally evaluated;
    \item \textit{Training data}: data samples used during model designing for fitting and optimizing the learning algorithm.
\end{itemize}

The test sets comprehend five pieces of data reserved through random subsampling without replacement. At the beginning, we randomly reserve about 365 data samples for each test segment in a stratified way, which guarantees an equally distributed ratio of positive and negative classes throughout the test sets and the remaining piece of macro-training data. In this sense, we indirectly avoid solar max/min problems of flare rates.

Then, we must keep the test sets aside and do not use them again until the end, when we assess the output model prediction error, which refers to the generalization error over real unseen data \citep{Hastie:2009}. Since our aim is also to evaluate models under operational settings, test data can also be defined as simulated operational data.

After reserving the test sets, we split the remaining piece of macro-training data into 5-folds in a stratified cross-validated fashioned way, thus separating the five training sets from their corresponding validation segments. Splitting data in a 5-fold cross-validated manner is rather common in literature, since it helps to reduce the model variance and consequently leverage its performance \citep{Hastie:2009}. 

Similar to the test sets, which we reserve and only come back at the end, we also reserve the validation sets and only use them to assess the validation error prior to the model evaluation over the test sets. The purpose of validation data is similar to that of test data; however, validation samples cannot be labeled operational (i.e., they are used for the final decision-making), but they are still as pure as the test ones (by pure samples, we mean records that are not treated nor modified in any form during model designing). 

Finally, we use the five pieces of training data for fitting and optimizing the learning algorithm. We carry out several well-defined machine learning-based processes over each individual training set, namely: feature selection, hyper-parameters optimization, cost function analysis, and adjustment of cut-off point. Starting by feature selection (Figure~\ref{fig:feature-selection}), we shall further discuss all of them in the next subsections. 

The rationale for using a 5-fold-based splitting is solely related to our case study. Depending on the amount of available data or the impact of variance on features, other \mbox{$n$-fold} schema may be used. Provided that the number of desired data segments increase/decrease, validation and test pieces of data must be re-considered.

\subsection{Feature Selection}
\label{sec:feature-selection}
The aim of feature selection is to choose a representative and effective set of features for the output model. After splitting data into test, validation, and training sets, we evaluate a randomly initialized baseline learning algorithm over each training set using repeated randomized, stratified 5-fold cross-validation. In addition, we carry out some analyses of selecting and discarding features in each training segment. 

Designing accurate learning models is not usually an easy task, since right input data must be considered. Inputting noisy or useless features into learning systems can lead to poor system performance \citep{Han:2006}. 

Besides, a natural way to reduce a high data input dimensionality issue, which may decrease the learner's performance, is by discarding  weak features. Thus, feature selection is an imperative task to be performed to choose only the best attributes and discard those that do not significantly contribute to the predictive performance.

Concerning our methodology, we use two strategies to measure the usefulness of features and to assist in discarding some of them, namely a filter method and a wrapper schema \citep{Guyon:2003}. The former uses a proxy metric (i.e., some correlation-based measure) to score features and outputs a rank of their importance. Discarding features with a filter method alone is merely based on this rank, which is not our case.

On the other hand, the wrapper schema uses an ordinary learning model to score several distinct feature subsets. Choosing a specific set of features with a wrapper solely depends on the analysis of which set best optimized a specific evaluation score. In our study, we use a univariate feature selection method provided with the F-score measure as the filter~\citep{Bobra:2015} linked to a randomly initialized XGBoost algorithm as the baseline model.

\subsubsection{Univariate Feature Selection}
Univariate feature selection methods do not consider the correlation between features during their analyses. Instead, features are deemed independent, and their complimentary nature, on the other hand, are tested. An arbitrary feature alone may have poor performance when predicting a target; however, when combined with others, it can be stronger~\citep{Guyon:2003}. 

Therefore, complementary features do not necessarily share a correlated relationship, i.e., two independent features can be good predictors combined, but have no high correlation coefficient between them~\citep{Guyon:2003}. The literature on flare forecasting systems had already demonstrated several benefits of choosing features by univariate methods combined with the F-score measure, as posited in \cite{Bobra:2015}'s study. In Equation~\ref{eq:f-score} we show the F-score calculation~\citep{Chang:2008}:

\begin{equation}\label{eq:f-score}
F(i)=\frac{(\bar{x}_{i}^{+}-\bar{x}_{i})^{2}+(\bar{x}_{i}^{-}-\bar{x}_{i})^{2}}{\frac{1}{n^{+}-1}\sum_{k=1}^{n^{+}}(x_{k,i}^{+}-\bar{x}_{i})^{2}+\frac{1}{n^{-}-1}\sum_{k=1}^{n^{-}}(x_{k,i}^{-}-\bar{x}_{i})^{2}}
\end{equation}
\noindent where the inter-class variance is referred in the numerator; the variance sum within each class is referred in the denominator; $\bar{x}_{i}^{-}$ and $\bar{x}_{i}^{+}$ refer respectively to the average values of negative and positive samples; $\bar{x}_{i}$ is the feature average value; and $n^{-}$ and $n^{+}$ are the number of negative and positive samples respectively. 

\begin{figure*}[!htb]
\centering
\includegraphics[width=1.0\textwidth]{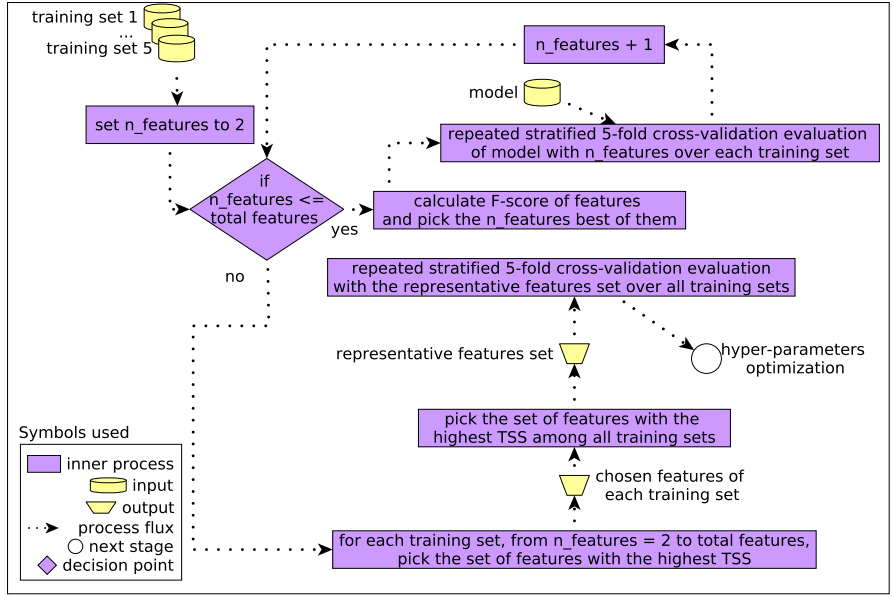}
\caption{Feature selection schema.}
\label{fig:feature-selection}
\end{figure*}

For each training segment (one at a time), we must calculate the F-score of the entire set of features and organize the output rank in descendant order, as Figure~\ref{fig:feature-selection} shows. In the beginning, we choose the two best features and evaluate them with the ensemble through repeated randomized, stratified 5-fold cross-validation. The number of the chosen best features is increased by one, and the corresponding true skill statistics (TSS) score~\citep{Jolliffe:2003} is subsequently recorded (see the Appendix Section A of this study for a concise reference for the TSS score). We repeat this process from the two best features to their total number.

This approach allows us choosing and reserving the five best feature sets at the end, i.e., the ones linked to the highest TSS training scores. Then, we reserve only the set of features associated with the highest TSS score among all training sets as the representative one. Subsequently, we re-evaluate the ensemble using this representative set over all training sets.

The rationale for using an optimization strategy based on TSS refers to how we calculate this metric. Surely, most scores could not individually represent two outcomes at once. However, one of the strengths of TSS is that it is calculated from two components that individually assess the success rates of the positive (TPR) and negative (TNR) outcome classes (see the Appendix B and C of this study for a concise reference for them).

Accordingly, increasing the TSS score naturally increases both TPR and TNR at once. Examples of researchers that optimized their models using TSS include \cite{Bobra:2015}, \cite{Bloomfield:2012}, and \cite{Nishizuka:2017}.

Furthermore, since TSS has a categorical nature and, at this point, we do not aim to deal with classifiers' decision thresholds, we keep models' thresholds at the default level (0.5) to calculate their corresponding confusion matrices. In later stages, the prediction thresholds shall be optimized, as described in Section~\ref{sec:cut-off-point-adjust}. Not only TSS, but also other scores may be optimized at this point (that depends on the requirements of the expected output forecast model)

\subsection{Optimization of Hyper-parameters}
\label{sec:hyper-parameters-optimization}
The aim of the optimization of hyper-parameters is to choose an affordable grid of parameters to adjust how the ensemble behaves and better leverages its generalization skills. After discarding some useless features, we must re-evaluate the ensemble on the training sets using appropriate parameters grids to seek the set of elements that best optimize the area under the receiver operating characteristic (ROC) curve (AUC) \citep{Witten:2011} (see the Appendix Section D of this study for a concise reference for AUC). 

Concerning the optimization of hyper-parameters theory, the aim is to fit a model $\mathcal{M}$ and reduce its loss function $\mathcal{L}(X^{(val)};\mathcal{M})$ in samples of validation data $X^{(val)}$ (examples of loss functions include the mean squared error for regression problems or the error rate for classification ones). The model $\mathcal{M}$ is thus trained by an algorithm $\mathcal{A}$ in samples of training data $X^{(tr)}$~\citep{Claesen:2015}. 

Overall, $\mathcal{A}$ has a set of hyper-parameters  $\lambda$, $\mathcal{M}=\mathcal{A}(X^{(tr)};\lambda)$, which is usually adjusted (optimized) to reduce $\mathcal{L}$. Besides reducing $\mathcal{L}$, $\mathcal{M}$ requires an affordable level of complexity, since complex models may poorly generalize unseen data (over-fitting). 

On the other hand, too simple models may not appropriately learn the data patterns (under-fitting). Accordingly, learning algorithms also provide their parameters $\lambda$ to allow the adjustment of their complexity.

Therefore, hyper-parameters can be used in a search process to seek a set $\lambda^{\star}$ that yield an optimal model, as Equation~\ref{eq:hyper-parameters-tuning} defines \citep{Claesen:2015}:

\begin{equation}\label{eq:hyper-parameters-tuning}
\lambda^{\star} = \lambda{\mathrm{arg \ min}} \  \mathcal{L}(X^{(val)};\mathcal{A}(X^{(tr)};\lambda)) = \lambda{\mathrm{arg \ min}} \ \mathcal{F}(\lambda; \mathcal{A},X^{(tr)},X^{(val)}, \mathcal{L})
\end{equation}

\noindent where the function $\mathcal{F}$ uses a set of hyper-parameters and outputs the loss value provided that $X^{(tr)}$, $X^{(val)}$ and $\mathcal{L}$ are given.

\subsubsection{Randomized Search for Hyper-parameters}
There are two common techniques usually used for searching hyper-parameters, namely the grid search and the random-based one. On the one hand, the former uses a grid of parameters with their corresponding data ranges and seeks all the possible combinations among them. On the other hand, the latter also uses this grid; however, the combinations are randomly sampled over the available ranges~\citep{Bergstra:2012}. 

Despite simple and effective, grid search can be costly in high dimensional input spaces. In turn, the random-based technique can be as effective as the former because not all parameters are equally important for adjustment and neither the input dimensions~\citep{Bergstra:2012}. In this sense, we use the random search along with the following XGBoost-based hyper-parameters grid: 

\begin{itemize}
    \item the number of inner decision trees;
    \item the maximum depth of the inner decision tress;
    \item the number of samples used to fit each inner tree (bagging strategy);
    \item the learning rate that drives how much each inner tree contribute to the loss gradient;
    \item the subsample of features used to fit each inner tree.
\end{itemize}

\begin{figure*}[!htb]
\centering
\includegraphics[width=1\textwidth]{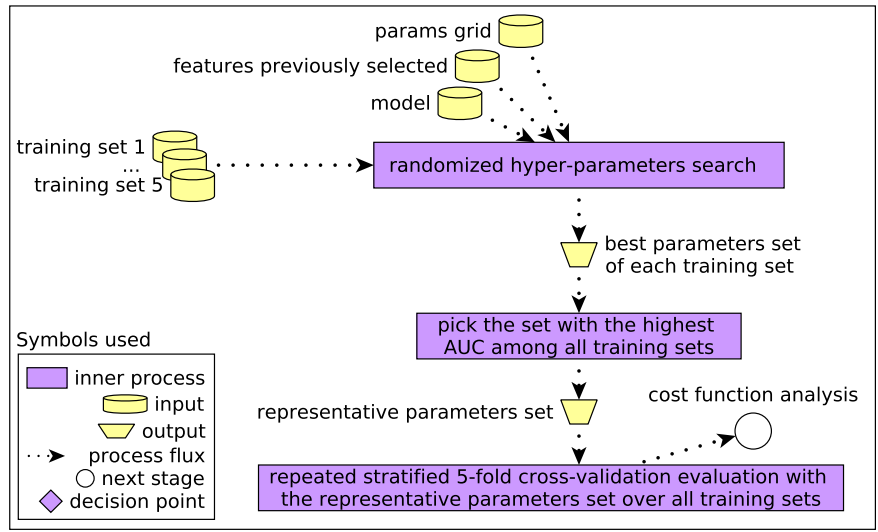}
\caption{Hyper-parameters optimization schema.}
\label{fig:hyper-parameters-optimization}
\end{figure*}

As we show in Figure~\ref{fig:hyper-parameters-optimization}, we perform a randomized search in each training set to search for the hyper-parameters set-up that most increase the AUC score. Furthermore, after the randomized searches, we must choose the representative set of parameters (the one that most increased the AUC among all training segments) and re-evaluate the baseline model provided with this set through each training set.

The AUC score represents the classifier skill for discriminating between positive and negative events. We propose AUC here instead of TSS since we want to guarantee that our model will have robustness for the probability of detecting false alarms at different prediction thresholds. In this sense, we seek a parameters set that does score high TPRs at increasing prediction thresholds without incurring high FPRs, thus leading to high AUCs. Not only AUC, but also other scores may be optimized at this point (that depends on the requirements of the expected output forecast model).

\subsection{Cost Function Analysis}
\label{sec:cost-function-analysis}

Besides choosing an affordable set of features and adjusting the ensemble complexity, we must also consider the imbalanced scenario of the class ratios with which we are dealing. Frequently, real binary data have imbalanced class ratios. Negative samples can be more common than the positive ones and vice versa. Before the initial splitting of our dataset, there were about 58\% of data referring to the positive class, whereas the rest represented the negative one (42\%).

Training models in imbalanced data may have a high cost and often produces over-fitted predictors, which cannot well generalize  minority samples. A straightforward way to address imbalanced data is to resample the training data using an appropriate technique (for instance, random over- or under-sampling) \citep{Batista:2004}. Other approaches involve changing the way classifiers report their performance, i.e., instead of only using the biased overall accuracy, reporting the class-specific performance metrics is useful -- which we have already done in this paper.

Over-sampling means randomly duplicating minority training samples to create a dataset that has as many negative as positive examples. However, this can lead to over-fitting in favor of the duplicated samples. On the other hand, under-sampling means randomly removing majority data samples. However, this can lead to important data loss \citep{Batista:2004}. 

Both techniques mentioned earlier -- and derivatives -- could be used to cope with imbalanced data in the methodology workflow. However, since we need an equally-distributed ratio of classes and do not want to harm our data distribution, a better way for dealing with imbalanced data is by cost sensitive learning \citep{Bobra:2015}. Instead of each class be simply correctly or incorrectly classified, a misclassification cost is assigned to each sample. In this sense, the classifier must then try to minimize the total misclassification cost~\citep{Elkan:2011}. That is the approach we use here, and involves an XGBoost algorithm functionality \citep{Chen:2016}, further explained in Section~\ref{sec:cost-sensitive-learning}.

\subsubsection{Cost-sensitive Learning}
\label{sec:cost-sensitive-learning}

The XGBoost algorithm has a feature for cost-sensitive learning, i.e., the learning approach that assigns distinct weights to the classes to adjust the class-specific hit rates. Thus, it provides an adjustment function that represents a weighted multiplication to every positive sample. Overall, this function represents the ratio of positive examples over the negative ones, $C_{p}/C_{n}$.

The cost-sensitive function of XGBoost assumes that positive events are rarer than the negative ones, and thus $C_{p}/C_{n}$ ratio values are expected to be lying around \big[1:100\big] by definition; however, since we have the opposite scenario (negative samples are rarer), we must work with the opposite interval of  $C_{p}/C_{n}$ ratio weights, \big[0.1:0.9\big]. In addition, it is worth mentioning that opposite weights solely refer to our case study and should be changed to the default ones as far as the positive class becomes rarer.

\begin{figure*}[!htb]
\centering
\includegraphics[width=1.0\textwidth]{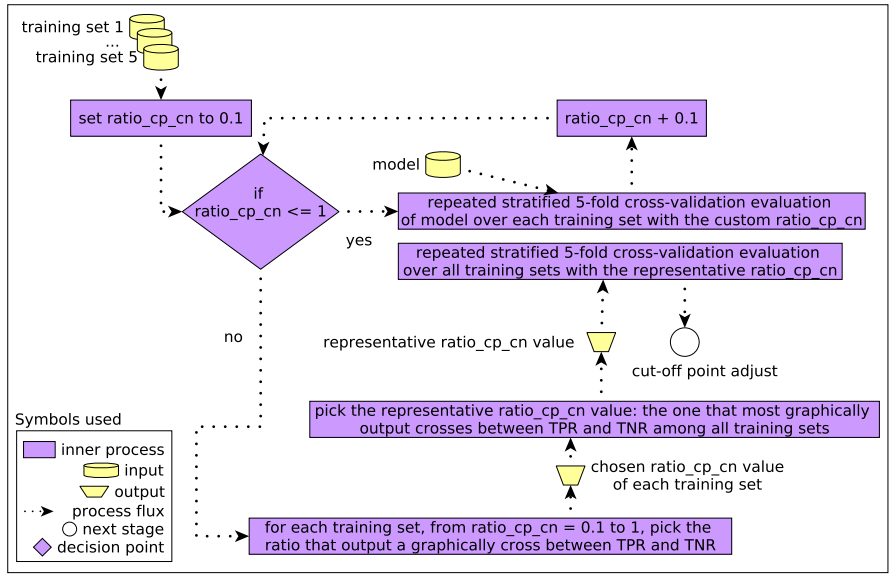}
\caption{Cost function analysis schema.}
\label{fig:cost-function-analysis}
\end{figure*}

\begin{figure*}[!htb]
\centering
\includegraphics[width=1.0\textwidth]{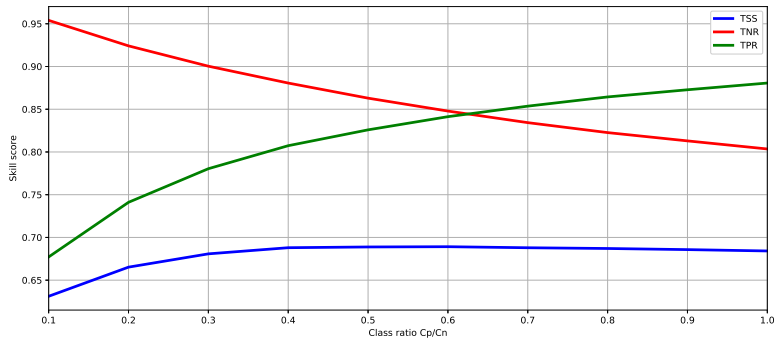}
\caption{Class ratio graphical plot for one of the training segments.}
\label{fig:class-ratio-plot}
\end{figure*}

As we show in Figure~\ref{fig:cost-function-analysis}, we seek the $C_{p}/C_{n}$ ratio of each training set (one at a time) which is linked to a graphical cross between TPR and TNR to keep both hit rates simultaneously at close levels. By a graphical cross between TPR vs. TNR, we mean the class ratio linked to $| TPR - TNR | \le 1$, as we show in Figure~\ref{fig:class-ratio-plot} for one of the training segments (from $C_{p}/C_{n} = 0.1$ to $C_{p}/C_{n} = 1$  at a 0.1 step-based increment, we repeat this process through all training sets). Then, we must choose the $C_{p}/C_{n}$ ratio that most produce graphical crosses as the representative one and re-evaluate the model provided by such through all the training sets.

Despite we may decrease the TPR somehow during this adjustment, the TNR will consequently increase. Since the TSS score is directly influenced by these two metrics (the higher the class-specific hit rates, the higher the TSS is), choosing the $C_{p}/C_{n}$ ratio that approximates the latter elements to a close level is an indirect way of increasing the former.

\subsection{Cut-off Point Adjustment}
\label{sec:cut-off-point-adjust}
Most learning algorithms output a posterior probability, and so does the XGBoost. This probability is turned into a binary positive prediction provided that a decision threshold $t$ is reached, i.e., usually, $t \geq 0.5$ outputs a positive answer. Also known as cut-off point of classifiers, $t = 0.5$ might not yield the best performance regarding the prediction of positive events and the produced false alarms.

Directly related to the TPR (recall), the positive predictive value (PPV) -- also known as precision~\citep{Zaki:2013} -- represents the accuracy for predicting positive events (see the Appendix Section F of this study for a concise reference for precision). Whereas the former accounts for the number of positive events correctly predicted, the latter indirectly refers to the skill to predict positive events.

Choosing certain cut-off point of classifiers directly affects both TPR and PPV. However, often, there will be two trade-offs between the class-specific recall and the precision scores. For instance, it would be rather easy to score a perfect positive recall (TPR = 1) by merely predicting all testing samples as positive, but the positive precision would be rather low \citep{Zaki:2013}. As a consequence, the number of false alarms would be highly increased.

On the other hand, the positive precision would be highly increased provided that we predict only a few testing samples as positive, i.e., those samples about which our model has the most confidence. Conversely, the positive recall would be highly decreased, and thus the number of false alarms would be highly decreased.

\begin{figure*}[!htb]
\centering
\includegraphics[width=1.0\textwidth]{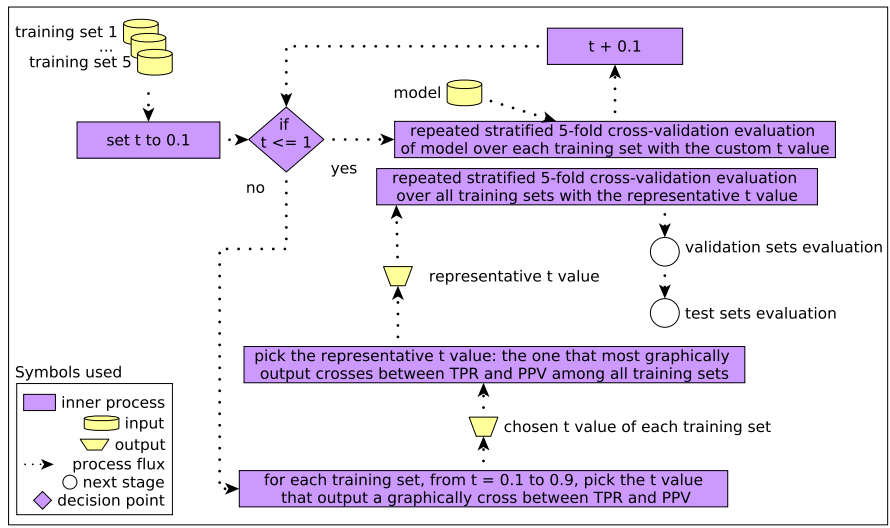}
\caption{Adjustment of the cut-off point schema.}
\label{fig:cut-off-point-adjust}
\end{figure*}

\begin{figure*}[!htb]
\centering
\includegraphics[width=1.0\textwidth]{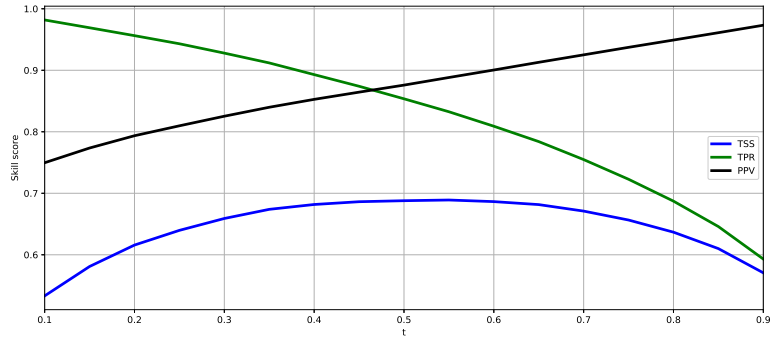}
\caption{Decision threshold graphical plot for one of the training segments.}
\label{fig:t-plot}
\end{figure*}

Ideally, an optimal balance cut-off point must be found in such a way that both recall and precision are simultaneously high~\citep{Zaki:2013}, and that is the aim of this process. According to Figure~\ref{fig:cut-off-point-adjust}, for each training set (one at a time), from $t = 0.1$ to $t = 0.9$ at a 0.1 step-based increment, we seek the $t$ value that output a graphical cross between precision and recall, i.e., the one linked to \mbox{$| TPR - PPV | \le 1$}, as we show in Figure~\ref{fig:t-plot} for one of the training segments. Then, we must choose the $t$ value that kept TPR closer to PPV over all training sets as the representative one. By using this rationale, we reach an optimal balance between recall and precision at the same time we keep the number of false alarms at low levels.

\subsection{Evaluation of Validation Sets}
\label{sec:validation-sets-evaluation}
After finding the optimal cut-off point for the classifier, we train five models of the same type (i.e., the same algorithm, features, hyper-parameters, $C_{p}/C_{n}$ ratio, and cut-off point) through each individual training set. We use these models to forecast their corresponding validation sets reserved at the beginning of the process.

Besides forecasting the validation sets with the output models, we also fit five baseline models over each training set and forecast their corresponding validation sets. Baseline models also have the same set-up; however, this set-up refers to the XGBoost algorithm during feature selection, noticeably a XGBoost model trained with randomly initialized parameters. 

We use these models to verify the methodology effectiveness in the final decision-making process: we verify if scores increased from the baseline models to the output ones and, if so, we proceed with the evaluation of the test sets.

\subsection{Evaluation of Test Sets}
\label{sec:test-sets-evaluation}
Last, but not least, provided that the predictive performance was increased during Section~\ref{sec:validation-sets-evaluation}, we can finally assess the generalization error over unseen data. Therefore, we use the previously trained baseline and output models to forecast their corresponding test sets.

\section{Results}
\label{sec:results} 

In this section, we will discuss the results (Table~\ref{tab:detailed-framework-results}) of carrying out the methodology as previously defined in Section~\ref{sec:methodology}. Besides TPR, TNR, PPV, TSS, and AUC, we will also provide other scores to allow a better understanding of models, namely negative predictive value (NPV) \citep{Zaki:2013}, overall accuracy (ACC) \citep{Han:2006}, and false alarm ratio (FAR) \citep{Jolliffe:2003}. We further describe these scores in the Appendix Sections F, G, and H of this study. In addition, to verify the significance of improvements reported between inner processes, we shall use a paired two-tailed Wilcoxon Signed-Rank Test \citep{Wilcoxon:1945}.

\begin{table}[!htb]
\centering
\caption{Detailed framework results.}
\label{tab:detailed-framework-results}
\resizebox{0.95\textwidth}{!}{
\begin{tabular}{cccccccccc}
\hline
\textit{Results}                                                                                                     & \textit{Forecasting Time} & \textit{ACC} & \textit{TPR} & \textit{TNR} & \textit{PPV} & \textit{NPV} & \textit{FAR} & \textit{AUC} & \textit{TSS} \\ \hline
\multirow{3}{*}{\begin{tabular}[c]{@{}c@{}}Feature\\ selection\end{tabular}}                                         & next 24 hours             & 0.69         & 0.71         & 0.67         & 0.73         & 0.60         & 0.20         & 0.71         & 0.38         \\
                                                                                                                     & 24-48 hours               & 0.68         & 0.74         & 0.60         & 0.70         & 0.55         & 0.23         & 0.69         & 0.33         \\
                                                                                                                     & 48-72 hours               & 0.67         & 0.68         & 0.65         & 0.73         & 0.58         & 0.22         & 0.69         & 0.34         \\ \hline
\multirow{3}{*}{\begin{tabular}[c]{@{}c@{}}Optimization of \\ hyper-parameters\end{tabular}}                         & next 24 hours             & 0.85         & 0.87         & 0.81         & 0.86         & 0.82         & 0.14         & 0.93         & 0.68         \\
                                                                                                                     & 24-48 hours               & 0.84         & 0.87         & 0.80         & 0.86         & 0.82         & 0.14         & 0.92         & 0.67         \\
                                                                                                                     & 48-72 hours               & 0.83         & 0.87         & 0.78         & 0.84         & 0.82         & 0.14         & 0.91         & 0.65         \\ \hline
\multirow{3}{*}{\begin{tabular}[c]{@{}c@{}}Cost function \\ analysis\end{tabular}}                                   & next 24 hours             & 0.85         & 0.85         & 0.85         & 0.88         & 0.80         & 0.12         & 0.93         & 0.69         \\
                                                                                                                     & 24-48 hours               & 0.84         & 0.85         & 0.84         & 0.88         & 0.80         & 0.12         & 0.92         & 0.68         \\
                                                                                                                     & 48-72 hours               & 0.83         & 0.83         & 0.82         & 0.87         & 0.78         & 0.13         & 0.91         & 0.66         \\ \hline
\multirow{3}{*}{\begin{tabular}[c]{@{}c@{}}Cut-off point \\ adjust\end{tabular}}                                     & next 24 hours             & 0.85         & 0.87         & 0.82         & 0.87         & 0.82         & 0.13         & 0.93         & 0.69         \\
                                                                                                                     & 24-48 hours               & 0.84         & 0.87         & 0.81         & 0.86         & 0.82         & 0.14         & 0.92         & 0.68         \\
                                                                                                                     & 48-72 hours               & 0.83         & 0.86         & 0.79         & 0.85         & 0.81         & 0.15         & 0.91         & 0.65         \\ \hline
\multirow{3}{*}{\begin{tabular}[c]{@{}c@{}}Evaluation of baseline \\ models through \\ validation sets\end{tabular}} & next 24 hours             & 0.66         & 0.63         & 0.69         & 0.83         & 0.51         & 0.17         & 0.68         & 0.33         \\
                                                                                                                     & 24-48 hours               & 0.77         & 0.79         & 0.75         & 0.81         & 0.73         & 0.19         & 0.79         & 0.54         \\
                                                                                                                     & 48-72 hours               & 0.61         & 0.38         & 0.91         & 0.72         & 0.56         & 0.08         & 0.66         & 0.30         \\ \hline
\multirow{3}{*}{\begin{tabular}[c]{@{}c@{}}Evaluation of output \\ models through \\ validation sets\end{tabular}}   & next 24 hours             & 0.85         & 0.86         & 0.83         & 0.87         & 0.81         & 0.13         & 0.93         & 0.69         \\
                                                                                                                     & 24-48 hours               & 0.84         & 0.87         & 0.81         & 0.86         & 0.82         & 0.14         & 0.92         & 0.68         \\
                                                                                                                     & 48-72 hours               & 0.83         & 0.86         & 0.79         & 0.85         & 0.81         & 0.15         & 0.92         & 0.65         \\ \hline
\multirow{3}{*}{\begin{tabular}[c]{@{}c@{}}Evaluation of baseline \\ models through \\ test sets\end{tabular}}       & next 24 hours             & 0.66         & 0.65         & 0.68         & 0.83         & 0.53         & 0.17         & 0.68         & 0.33         \\
                                                                                                                     & 24-48 hours               & 0.77         & 0.80         & 0.73         & 0.81         & 0.73         & 0.19         & 0.79         & 0.53         \\
                                                                                                                     & 48-72 hours               & 0.62         & 0.40         & 0.92         & 0.53         & 0.57         & 0.07         & 0.67         & 0.31         \\ \hline
\multirow{3}{*}{\begin{tabular}[c]{@{}c@{}}Evaluation of output \\ models through \\ test sets\end{tabular}}         & next 24 hours             & 0.86         & 0.89         & 0.81         & 0.87         & 0.84         & 0.13         & 0.94         & 0.70         \\
                                                                                                                     & 24-48 hours               & 0.84         & 0.87         & 0.79         & 0.85         & 0.82         & 0.15         & 0.92         & 0.67         \\
                                                                                                                     & 48-72 hours               & 0.85         & 0.87         & 0.81         & 0.86         & 0.82         & 0.14         & 0.92         & 0.68         \\ \hline
\end{tabular}}
\end{table}

\subsection{Results from Feature Selection}
Overall, during feature selection, we observed low TSS scores over all forecasting horizons, with their values respectively lying around \big[0.33:0.38\big]. With the ACC scores, in turn, we had just regular results, and so we did with the TNR and TPR ones. Although the TPR score has achieved results in the interval \big[0.68:0.74\big], the TNR only lied around \big[0.60:0.67\big].

Concerning the positive and negative precision scores, they also achieved just regular results. Whereas the PPV lied around \big[0.70:0.73\big], no NPV result surpassed 0.60.


Although our models did score AUC results close to 0.70, their decreases of about 0.3 on average pointed that false alarms somehow occurred at high frequency. FAR scores corroborate this finding for confirming that false alarms were observed in at least 20\% of the times (next 24 hours).

At this point, it is worth commenting that there were features with similar natures selected for the classifiers. For instance, the univariate analysis selected features comprehending almost the entire sliding time windows of the sunspot number and radio flux with high F-scores over all time horizons -- except the number of sunspots of 4 days before the $t$ instant for 24-48 hrs (numbers preceded by colons refer to F-scores of features):

\begin{itemize}
    \item next 24 hr
        \begin{itemize}
            \item radio\_flux\_10.7cm\_t4 : 2514.79
            \item radio\_flux\_10.7cm\_t3 : 2676.29
            \item radio\_flux\_10.7cm\_t2 : 2856.75
            \item radio\_flux\_10.7cm\_t1 : 3023.52
            \item radio\_flux\_10.7cm\_t : 3191.59
            \item sesc\_sunspot\_number\_t4 : 2243.35
            \item sesc\_sunspot\_number\_t3 : 2432.46
            \item sesc\_sunspot\_number\_t2 : 2626.52
            \item sesc\_sunspot\_number\_t1 : 2796.06
            \item sesc\_sunspot\_number\_t : 2942.71
        \end{itemize}
    \item 24-48 hrs
        \begin{itemize}
            \item radio\_flux\_10.7cm\_t4 : 2443.53 
            \item radio\_flux\_10.7cm\_t3 : 2578.98 
            \item radio\_flux\_10.7cm\_t2 : 2734.72 
            \item radio\_flux\_10.7cm\_t1 : 2914.88 
            \item radio\_flux\_10.7cm\_t : 3110.54 
            \item sesc\_sunspot\_number\_t3 : 2167.06 
            \item sesc\_sunspot\_number\_t2 : 2358.84 
            \item sesc\_sunspot\_number\_t1 : 2536.64 
            \item sesc\_sunspot\_number\_t : 2728.36
        \end{itemize}
    \item 48-72 hrs
        \begin{itemize}
            \item radio\_flux\_10.7cm\_t4 : 2205.92
            \item radio\_flux\_10.7cm\_t3 : 2335.99
            \item radio\_flux\_10.7cm\_t2 : 2473.10
            \item radio\_flux\_10.7cm\_t1 : 2626.34
            \item radio\_flux\_10.7cm\_t : 2787.47
            \item sesc\_sunspot\_number\_t4 : 1981.64
            \item sesc\_sunspot\_number\_t3 : 2112.54
            \item sesc\_sunspot\_number\_t2 : 2263.49
            \item sesc\_sunspot\_number\_t1 : 2432.12
            \item sesc\_sunspot\_number\_t : 2584.73
        \end{itemize}
\end{itemize}

We are aware that the nature of some features of ours may be similar (correlated features), which can somehow directly lead to redundancy in the input of classifiers (features colinearity). However, we are using a method to test the complementary nature of features instead of their correlation -- that is, we aimed at identifying independent features that led to good predictors when combined regardless of their correlation.

\subsection{Results from Optimization of Hyper-parameters}
Overall, the search for hyper-parameters led to positive effects in all the scores. Except the FAR, which we positively decreased by 0.065 (next 24, $p < 0.1$), 0.086 (24-48, $p < 0.1$), and 0.061 (48-72 hours, $p < 0.1$), we increased the other scores, such as the TPR and TNR -- in this case, $p$ refers to the $p-value$ calculated with the Wilcoxon Signed-Rank Test. Directly related to class-specific recalls, the TSS had the following improvements: 0.308 (next 24, $p < 0.1$), 0.339 (24-48, $p < 0.1$), and 0.31 (48-72 hours, $p < 0.1$). 

Since the $p-values$ mentioned earlier were less than our defined confidence interval ($\alpha = 0.1$), we could successfully reject our Wilcoxon Signed-Rank Test $H0$ hypothesis (the difference of two paired dependent observations equals zero). This means that the chance of type-1 statistical error (i.e., rejecting a correct H0) is small (10\%). Therefore, we could consider the average values of TSS and FAR between feature selection and hyper-parameters optimization not equal than our defined expected difference (zero), thus turning the differences between the average values statistically significant.

Furthermore, PPV and NPV also increased. The positive precision increased by 0.115 ($p < 0.1$) in the worst scenario, 48-72 hours, whereas the negative one reached a score of 0.226 points higher than feature selection ($p < 0.1$) also in the worst scenario, next 24 hours.

Besides the decreased number of predicted false alarms, the probability of detecting such events also decreased, thus increasing AUC. Since we had high AUC scores over all scenarios, we state that we had low FPR scores along with high TPR ones for increasing prediction thresholds. In this sense, the AUC score increased by 0.224 (next 24, $p < 0.1$), 0.235 (24-48, $p < 0.1$), and 0.221 (48-72 hours, $p < 0.1$).

Despite the positive results achieved at this point, there should be further improvements. For instance, the higher positive hit rates when compared with the negative ones suggest slightly imbalanced scenarios of class ratios, which will be addressed in the analysis of the classifier cost-functions. In addition, there is also need for finding an optimal cut-off point for classifiers, provided that we have now robust AUC scores.


\subsection{Results from Cost Function Analysis}
Overall, we had subtle -- but positive -- results with the cost function analysis. Two scores remained at the same level, namely the ACC and AUC. Others, in their turn, had positive increases such as TNR and PPV.

Observed trend inversions of scores between TPR vs. PPV and TNR vs. NPV were directly related to the class-specific precision-recall trade-offs. We expected these inversions, since we wanted to balance (approximate) the class-specific hit rates by penalizing individual outcome classes.

Furthermore, we had positive effects with FAR and TSS. Whereas the former positively decreased by 0.018 (next 24, $p < 0.1$), 0.019 (24-48, $p < 0.1$), and 0.022 (48-72 hours, $p < 0.1$), the latter increased by 0.01 over all forecasting horizons ($p < 0.1$). 

The aforementioned results are closely related to each other and may also be used to explain the TSS increase along with the TPR decrease. The FAR may have been decreased in response to a decreasing FPR, thus leading to a TSS increase despite the TPR decrease. Accordingly, TSS is also represented by $TPR-FPR$, with the FPR defined as $1-TNR$.


\subsection{Results from Cut-off Point Adjustment}
During the cut-off point adjustment, we managed to keep ACC, AUC, and TSS at the same level of the cost function analysis (except the TSS for 48-72 hours, which decreased by 0.01, $p < 0.1$). In addition, there were some expected inversions between TPR vs. PPV and TNR vs. NPV, since we moved the cut-off points of classifiers. A a result, we increased the TPR scores and slightly decreased the PPV ones. 

At this point, classifiers are ready for the final decision-making process (evaluation of validation sets), i.e., we will assess the performance of classifiers over data that were not directly used during model designing. If models output a reliable performance, we shall proceed with the evaluation of test sets (unseen data).


\subsection{Evaluation of Validation Sets}
Overall, during the evaluation of validation sets, we observed increases of 0.361 (next 24, $p < 0.1$), 0.138 (24-48, $p < 0.1$), and 0.357 (48-72 hours, $p < 0.1$) with TSS in comparison to the baseline models. Not only TSS, but also AUC increased (all models scored at least 0.92, $p < 0.1$). Those results confirmed the effectiveness of models and made us proceed to forecast truly unseen data.

\subsection{Evaluation of Test Sets}
Similar to what we observed during the evaluation of validation sets, results also improved from the baseline to the output models here. Not only we were able to improve TSS by 0.368 (next 24, $p < 0.1$), 0.133 (24-48, $p < 0.1$), and 0.357 (48-72 hours, $p < 0.1$), but also managed to increase AUC over all forecasting horizons ($p < 0.1$), thus consequently maintaining the false alarms under control with low FAR scores. Those results refer to the forecast performance as if systems truly did forecast new data.
\section{Literature Analysis}
\label{sec:literature-analysis} 
We shall use this section to position our forecast performance within the specialized literature, namely the studies by \cite{Muranushi:2015}, \cite{Huang:2018}, \cite{Ahmed:2013}, \cite{Yang:2013}, \cite{Domijan:2019}, \cite{Nishizuka:2018}, \cite{Hada-Muranushi:2016}, \cite{Colak:2009}, \cite{Leka:2018}, \cite{Bloomfield:2012}, and \cite{Crown:2012}.

As argued by \cite{Barnes:2008} and \cite{Barnes:2016}, directly comparing scores from different methods may be meaningless because of the several underlying differences of involved systems, such as the data segments used for training and testing, the type of prediction (full-disk or AR-by-AR), and how flare events were defined. It does not become evident if some forecasting merit led to higher scores or if differences in results are due to data that differed.

Therefore, numbers shown in this section should not be considered direct head-to-head comparisons between different papers. Moreover, one should not conclude that a model had some score improved/decreased against a reference. 

In this sense, we shall drive this section to group systems by their TSS scores. As argued by \cite{Bloomfield:2012}, TSS can be roughly used as a reference between different forecast approaches. Inside the TSS groups, we will analyze systems regarding the identified interplay effects between accuracy, recall, precision, and the number of false alarms, seeking over-fitted forecasts. 

Table~\ref{tab:results-comparison} presents the forecast performance of the models mentioned earlier, along with our models. Overall, we included models considering the following criteria:

\begin{itemize}
\item Systems designed to forecast events in the next 24, 24-48, and 48-72 hours. Overlapping forecasting horizons were not be included, i.e., events in the next 48 or 72 hours.
\item Systems able to forecast $\geq C$ class flares.
\item Systems that output full-disk or active-region-based forecasts.
\item Systems designed to rely on data of mixed nature, i.e., input features based on vector magnetic data or on the magnetic configuration of sunspots.
\item Literature focusing on the last 10 years. We did not want to disregard literature of longer periods, but our focus was to provide an in-depth look at the state-of-the-art, instead.
\item In addition to system forecasts, we also included human-based ones, made at prediction centers.

\end{itemize}

Besides, as Table ~\ref{tab:results-comparison} shows, we also managed to distinguish systems between two different groups: biased results or not. Biased results refer to systems that we identified some bias in the reported performance. The two criteria we used to classify systems as biased included: (i) evaluation without truly unseen data and (ii) use of magnetograms with ARs near the solar disk center (within a defined radius). 

Concerning the first criterion, reserving unseen data during systems design gives a true performance estimate while distinguishing between flare and non-flare events. As an example, \cite{Ahmed:2013} proposed to split their datasets into training and testing data yearly. In our case, we designed the methodology with the test data splitting right at the beginning on a random-based fashioned way.

On the other hand, we also classified as biased the studies whose authors only used magnetograms with ARs near the solar disk center. As stated by~\cite{Nishizuka:2017}, choosing such samples of ARs causes uncertainty about the reported scores, thus weakening the interpretation of the results for operational purposes, since in real operational environments the systems must behave with ARs at any location in the solar disk. Henceforth, let the following papers be classified as biased for having used magnetograms with ARs within a defined radius from the disk center: \cite{Muranushi:2015} (\ang{69}), \cite{Huang:2018} (\ang{30}), \cite{Ahmed:2013} (\ang{45}), \cite{Yang:2013} (\ang{30}), and \cite{Domijan:2019} (\ang{45}). 

As an example, \cite{Falconer:2014} designed their forecast approach with data from ARs near the disk center (\ang{30}). However, when deployed it into a real forecasting environment, their system warned reduced accuracy for predictions beyond \ang{30}. Nevertheless, biased results are not wrong, but they refer to certain evaluation conditions that can mask the real generalization skill of models when forecasting new data of any type. 

For coherence, unbiased systems took our test results as a reference. On the other hand, we took our validation results as a reference for biased systems. The bias of those results relates to our validation samples, which were indirectly used for decision-making when intersecting between the inner processes, i.e., the cross-cutting decisions for choosing features, hyper-parameters, $C_{p}/C_{n}$ ratios, and decision thresholds. 

\begin{table*}
\centering
\caption{Literature state-of-the-art.}
\label{tab:results-comparison}
\resizebox{\textwidth}{!}{
\begin{threeparttable}
\centering
\begin{tabular}{cccccccc}
\hline
\textit{Forecasting Time} & \textit{Authorship}      & \textit{Biased results} & \textit{ACC} & \textit{TPR} & \textit{TNR} & \textit{TSS} & \textit{FAR} \\ \hline
next 24 & xgboost\tnote{a} & no & 0.85 & 0.86 & 0.83 & 0.69 & 0.13         \\
24-48 & xgboost\tnote{a} & no & 0.84 & 0.87 & 0.81 & 0.68 & 0.14        \\
48-72 & xgboost\tnote{a} & no & 0.83 & 0.86 & 0.79 & 0.65 & 0.15        \\
next 24 & \cite{Muranushi:2015}\tnote{b} & no & 0.81 & 0.80 & 0.83 & 0.63 & 0.07 \\
next 24 & \cite{Huang:2018}\tnote{c} & no & 0.76 & 0.73 & 0.76 & 0.49 & 0.65 \\
next 24 & \cite{Ahmed:2013}\tnote{d} & no & 0.96 & 0.52 & 0.98 & 0.53 & 0.25 \\
next 24 & \cite{Yang:2013}\tnote{e} & no & 0.76 & 0.61 & 0.84 & 0.47 & - \\
next 24 & \cite{Domijan:2019}\tnote{f} & no & 0.89 & 0.95 & 0.89 & 0.84 & - \\
next 24 & \cite{Domijan:2019}\tnote{g} & no & 0.81 & 0.87 & 0.80 & 0.67 & - \\
\hline
next 24 & xgboost\tnote{a} & yes & 0.86 & 0.89 & 0.81 & 0.70 & 0.13      \\
24-48 & xgboost\tnote{a} & yes & 0.84 & 0.87 & 0.79 & 0.67 & 0.15         \\
48-72 & xgboost\tnote{a} & yes & 0.85 & 0.87 & 0.81 & 0.68 & 0.14         \\
next 24 & \cite{Nishizuka:2018}\tnote{h} & yes & 0.82 & 0.81 & 0.82 & 0.63 & 0.47 \\
next 24 & \cite{Hada-Muranushi:2016}\tnote{i} & yes & 0.66 & 0.72 & 0.57 & 0.30 & 0.30 \\
next 24 & \cite{Crown:2012}\tnote{j} & yes & 0.90 & 0.63 & 0.94 & 0.57 & 0.40 \\
next 24 & \cite{Colak:2009}\tnote{k} & yes & 0.81 & 0 .81 & - & - & 0.30 \\
next 24 & \cite{Leka:2018}\tnote{l} & yes & 0.75 & 0.69 & 0.82 & 0.51 & 0.17 \\
24-48 & \cite{Leka:2018}\tnote{l} & yes & 0.77 & 0.71 & 0.83 & 0.55 & 0.16 \\
48-72 & \cite{Leka:2018}\tnote{l} & yes & 0.71 & 0.60 & 0.85 & 0.45 & 0.17 \\
next 24 & \cite{Leka:2018}\tnote{m} & yes & 0.92 & 0.30 & 0.99 & 0.29 & 0.30 \\
24-48 & \cite{Leka:2018}\tnote{m} & yes & 0.93 & 0.27 & 0.99 & 0.26 & 0.25 \\
48-72 & \cite{Leka:2018}\tnote{m} & yes & 0.94	& 0.27 & 0.99 & 0.26 & 0.30 \\
next 24 & \cite{Bloomfield:2012}\tnote{n} & yes & 0.71 & 0.75 & 0.70 & 0.45 & 0.64 \\
\bottomrule 
\end{tabular}

\begin{tablenotes}
\item[a] Our model.
\item[b] Scores calculated over the confusion matrix of Table 5~\citep{Muranushi:2015}. 
\item[c] Scores calculated over the confusion matrix of Table 4~\citep{Huang:2018}.   
\item[d] Scores collected from Table 6~\citep{Ahmed:2013} (training and testing columns: operational set up). TSS calculated over TPR and TNR.
\item[e] Scores collected from Table 4~\citep{Yang:2013}.
\item[f] Scores collected from Table 4~\citep{Domijan:2019} (p = 0.05). 
\item[g] Scores collected from Table 7~\citep{Domijan:2019} (p = 0.20). 
\item[h] Scores calculated over the confusion matrix of Figure 5~\citep{Nishizuka:2018}.
\item[i] Scores calculated over the confusion matrix of Table 5~\citep{Hada-Muranushi:2016}. 
\item[j] Scores calculated over the confusion matrix of Table 4~\citep{Crown:2012}.
\item[k] Scores collected from Table 3~\citep{Colak:2009}.
\item[l] Scores calculated over the confusion matrix of Fig. 13~\citep{Leka:2018} (full-disk forecasts). 
\item[m] Scores calculated over the confusion matrix of Fig. 13~\citep{Leka:2018} (active-region-based forecasts). 
\item[n] Scores collected from Table 4~\citep{Bloomfield:2012}. TNR calculated over TPR and TSS.

\end{tablenotes}

\end{threeparttable}
}
\end{table*}

\subsection{Remarks from the Biased Results Analysis}

The first group of papers we shall discuss scored \mbox{$0.65 \leq TSS < 0.85$}, namely the researches by \cite{Domijan:2019} (TSS = 0.67 and 0.84) and our models (TSS = 0.69, 0.68, and 0.65). By achieving TSS scores in this interval, they were able to keep their TPR and TNR scores at close levels. Noticeably, their TPRs ranged on {\big[0.86:0.95\big]}, whereas their TNRs lied around {\big[0.79:0.89\big]}. Since they scored ACCs ({\big[0.81:0.89\big]}) close to the TPR and TNR intervals, we could not suggest over-fitted systems (i.e., high ACCs and TNRs with low TPRs).

On the other hand, the second group scored \mbox{$0.45 \leq TSS < 0.65$}, namely the researches by \cite{Yang:2013} (TSS = 0.47), \cite{Huang:2018} (TSS = 0.49), \cite{Ahmed:2013} (TSS = 0.53), and \cite{Muranushi:2015} (TSS = 0.63). Despite having scored TSSs at lower levels, they scored TPRs and TNRs most of the time without preference for a specific class. Whereas their positive recall ranged on {\big[0.52:0.80\big]}, the negative one varied over {\big[0.76:0.98\big]}.

However, except \cite{Yang:2013}, which did not inform their predicted number of false alarms or positive precision, some papers somehow produced high FARs, such as \cite{Ahmed:2013} (FAR = 0.25) and \cite{Huang:2018} (FAR = 0.65). What we argue here is that by trying to improve TPR, those authors may have harmed the positive precision of their systems (i.e., the precision-recall trade-off), thus increasing the number of predicted false alarms. \cite{Huang:2018}'s precision corroborates this statement, since their score only equaled 0.35.

In addition, we could identify \cite{Ahmed:2013}'s model as a potential over-fitted system in favor of the positive class. They scored high ACC (0.96) along with TNR (0.98), but their TPR only equaled 0.52. Thus, their model was not able to generalize well positive samples of their data.

\subsection{Remarks from the Unbiased Results Analysis}

Except \cite{Colak:2009}, which did not provide their TSS, the first group of papers we could infer scored \mbox{$0.50 \leq TSS < 0.75$}, namely the researches by \cite{Leka:2018} (TSS = 0.51 and 0.55), \cite{Nishizuka:2018} (TSS = 0.63), and our models (TSS = 0.67, 0.68, and 0.7). Once more, we could not suggest over-fitted systems, since the TPR scores ranged over {\big[0.69:0.89\big]} and the TNRs {\big[0.79:0.83\big]}, with ACCs varying over {\big[0.77:0.86\big]}. However, we could observe some harmed precision scores, which somehow leveraged the number of false alarms, such as with \cite{Nishizuka:2018} (FAR = 0.64). In this case, \cite{Nishizuka:2018}'s PPV corroborates their high FAR number, for having equaled only 0.53.

On the other hand, the rest of papers scored \mbox{$0.25 \leq TSS < 0.50$}, namely the researches by \cite{Leka:2018} (TSS = 0.26, 0.29, and 0.45), \cite{Hada-Muranushi:2016} (TSS = 0.30), and \cite{Bloomfield:2012} (TSS = 0.45). Within this group, the TPR scores varied over \big[0.27:0.72\big], whereas the TNRs \big[0.57:0.99\big]. As ACCs lied around higher levels (\big[0.66:0.94\big]), we could infer some over-fitted systems in favor of the positive class, such as the active region based forecasts by \cite{Leka:2018} (they had high TNRs and ACCs, along with low TPRs).

In addition, we could also observe some cases of FAR $\geq 0.30$, such as with \cite{Hada-Muranushi:2016} (FAR = 0.3) and \cite{Bloomfield:2012} (FAR = 0.64). For having scored a positive precision of 0.7, we could suggest the precision-recall trade-off effects for \cite{Hada-Muranushi:2016}'s paper too. \cite{Bloomfield:2012}, in turn, did not provide their PPV.

Last, but not least, we can see the forecasting horizons of the previous papers noteworthily differing. For instance, whereas \cite{Leka:2018}, \cite{Hada-Muranushi:2016}, \cite{Colak:2009}, \cite{Bloomfield:2012}, and \cite{Nishizuka:2018} designed systems to forecast events within the next 24 hours, \cite{Leka:2018} also considered forecasting flares in 24-48 and 48-72 hours ahead along with our models. Therefore, the scores mentioned earlier should not be by any means directly judged one against the other. Different underlying event-day climatology resulting from the chosen forecasting horizon can lead to significant differences in forecast performance, even for a system being applied unmodified over two different forecasting horizons.

\section{Conclusions} 
\label{sec:conclusion}
Aiming at proposing some standardization to support the design of flare forecasting systems, we suggested a methodology to cope with most of the aspects with which the literature is concerned while developing such models. To validate our methodology, we assembled a dataset based on daily aggregated solar data provided by the NOAA/SWPC, including measures of the radio flux, x-ray background flux, sunspot number and area, and the weighted mean flare rate of magnetic and McIntosh classes. We aimed at forecasting solar flares $\geq C$ class up to 3 days ahead, namely forecasting them in the next 24, 24-48, and 48-72 hours. We also designed our input predictive features through a sliding time window to represent how our data evolved throughout five days.

Concerning our methodology effectiveness, the extreme gradient boosting tree models, designed under our pipeline, increased their operational TSS by 0.37 (next 24, $p < 0.1$), 0.13 (24-48, $p < 0.1$), and 0.36 (48-72 hours, $p < 0.1$) while also increasing the AUC scores, which have augmented from 0.68, 0.79, and 0.67 to 0.94 ($p < 0.1$), 0.92 ($p < 0.1$), and 0.92 ($p < 0.1$) in the next 24, 24-48, and 48-72 hours, respectively. Those increases refer to comparisons with random baseline models established at the beginning of the methodology.

Not only could we increase TSS and AUC, but we also did not manage to harm classifiers' precision. In this sense, the number of predicted false alarms has not been severely affected, thus positively decreasing the FAR score over all forecasting horizons.

Moreover, it is worth mentioning that we believe our models could also adhere to hybrid prediction schemas, such as the approach by NOAA/SWPC, according to which an expert system estimates the predictions at first and then human experts adjust them \citep{Crown:2012}. As argued by \cite{Murray:2017}, this would be helpful to reduce some climatology effects that would certainly affect predictions made over longer forecasting horizons (for instance, which ARs may leave or return to the solar disk within the next few days or how the ARs evolve while crossing the disk).


Furthermore, for considering the inner process as well-defined boxes with input and outputs, several other machine learning optimization processes could be easily integrated into the methodology, enabling not only the integration of new processes, but also the adjustment of those currently existing. The list of custom aspects of the methodology includes, but is not restricted to:

\begin{enumerate}[(i)]
\item Event type or magnitude: we set the event definition to represent $\geq C$ class flares in our case study. However, depending on the used input features, the event magnitude or type could be easily adjusted. For instance, the methodology could be used along with the forecasting of specific classes of flares or other events such as coronal mass ejections, and mass and speed of solar energetic particles. This aspect grants the methodology a wider application in solar weather research.
\item Feature selection method: although we did focus our feature selection method on a univariate schema provided with the F-score, other algorithms would be easily joined in the methodology pipeline, such as the Pearson correlation analysis.
\item Inner validation scores: TSS and AUC should not be considered the only options for optimizing through the inner processes. The cross-cutting decisions among inner processes can comprise the analysis of the F1-score \citep{Han:2006}, Heide Skill Score (HSS) \citep{Jolliffe:2003}, among others.
\item Search of hyper-parameters: the random-based approach for seeking the representative set of hyper-parameters could be replaced with the grid-based one, for instance.
\item Algorithm type: the proposed pipeline is not restricted to the case study models, since algorithms are treated as black boxes. Thus, provided that new algorithms are inputted with their corresponding parameter grids, new models would be designed, trained, and evaluated with few or no efforts.
\item Imbalanced data treatment method: the pipeline is not restricted to the cost function analysis for dealing with imbalanced data. Instead, some data re-sampling approaches could be used, such as the Synthetic Minority Over-Sampling Technique (SMOTE)~\citep{Chawla:2002}.
\end{enumerate}

As future research, we intend to expand our methodology focusing on adjusting its techniques to improve the performance of the output models and to deploy them, as well as the methodology, into a real operational space weather prediction engine. Since we based our dataset on daily solar data and sunspot region summaries, we can think of an operational system adhering to the data availability provided by NOAA. 

Data from DSD are always issued at 02:30, 08:30, 14:30, and 20:30 UTC \linebreak \citep{Noaa:2011}, whereas the SRS repository is updated at 00:30 UTC \linebreak \citep{Noaa:2008}. Since SRS files are issued at 00:30 and refer to previous days, it’s worth considering this time as a reference for daily forecasts (with data from DSD previously issued at 20:30). At each execution, the models will collect the last five days of available DSD records, integrate with their corresponding SRS data, and forecast the $\geq C$ class flares existence up to three days ahead.

Besides deploying the methodology into a real-time forecasting environment to generate forecast models on-demand for further experiments, we also intend to make the methodology source code available to the community in GitHub as soon as we finish its code documentation. Opening our source code will let other researchers improve and use it with their own datasets. 

%
\appendix 

In this paper we used some well-defined scores according to the proposed methodology to design the forecasting models. In this Appendix Section we present all of them, namely the true skill statistics (TSS) \citep{Jolliffe:2003}, true positive rate (TPR) \citep{Han:2006}, true negative rate (TNR) \citep{Han:2006}, area under curve (AUC) \citep{Witten:2011}, false positive rate (FPR) \citep{Witten:2011}, positive predictive value (PPV) \citep{Zaki:2013}, negative predictive value (NPV) \citep{Zaki:2013}, overall accuracy (ACC) \citep{Han:2006}, and false alarm ratio (FAR) \citep{Jolliffe:2003}.

\section{TSS}
The TSS score ranks a model performance over a scale lying on the interval \big[-1:1\big], according to which results close to $-1$ mean all incorrect predictions and those close to $1$ mean all correct predictions. The TSS was proposed as a quality measure that combines both class-specific hit rates without being affected by imbalanced data scenarios~\citep{Bloomfield:2012}. In Equation~\ref{eq:tss} we show how to calculate this score:

\begin{equation}\label{eq:tss}
TSS = TPR + TNR - 1 
\end{equation}
\noindent where the TPR refers to the true positive rate and the TNR is the true negative rate.

\section{TPR}
The TPR -- also known as the positive recall or probability of detection (POD) -- accounts for the number of positive samples correctly predicted \citep{Han:2006}. In Equation~\ref{eq:tpr} we show how to calculate the TPR:

\begin{equation}\label{eq:tpr}
TPR = \frac{TP}{TP+FN} 
\end{equation}
\noindent where the TP accounts for the true positives (positive samples predicted as positive) and the FN are the false negatives (positive samples predicted as negative) \citep{Han:2006}. The TPR lies around the interval \big[0:1\big], in which higher values are better.

\section{TNR}
The TNR refers to the true negative rate -- also known as the negative recall -- and accounts for the number of negative samples correctly predicted~\citep{Han:2006}. Similarly to the TPR, the TNR is scored in the same interval, as we show in Equation~\ref{eq:tnr}:

\begin{equation}\label{eq:tnr}
TNR = \frac{TN}{TN+FP} 
\end{equation}
\noindent where the TN accounts for the true negatives (negative samples predicted as negative) and the FP are the false positives (negative samples predicted as positive) \citep{Han:2006}.

\section{AUC}
\begin{figure*}[!htb]
\centering
\includegraphics[width=1.0\textwidth]{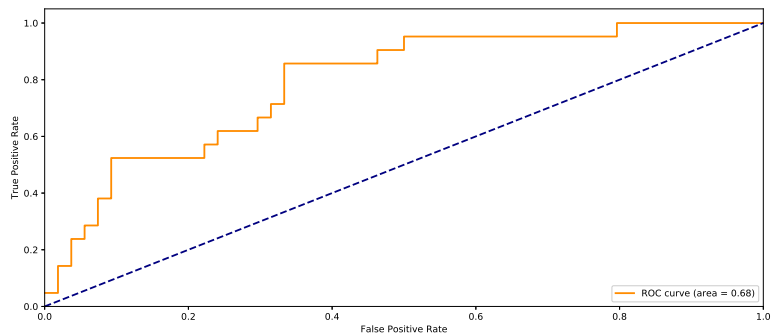}
\caption{Example of ROC curve graphical plot.}
\label{fig:roc-plot}
\end{figure*}

The AUC measures the two-dimensional area underneath the receiver operating characteristic (ROC) curve, which graphically analyzes the TPR score (y-axis) against the false positive rate (FPR) (x-axis) for a set of increasing probability thresholds to make the yes/no decisions, i.e., 0.1, 0.2, 0.3 etc (Figure~\ref{fig:roc-plot}). Also known as the probability of a false detection (POFD) or the false alarm rate, the FPR calculates the probability of detecting false alarms among the negative predictions, as we show in Equation~\ref{eq:fpr}~\citep{Witten:2011}:

\begin{equation}\label{eq:fpr}
FPR = \frac{FP}{TN + FP} 
\end{equation}
\noindent where the FP are the false positives and the TN are the true negatives. 

Best classifiers score the AUC next to the graph left-hand corner (FPR = 0 and TPR = 1). On the other hand,  worst classifiers score next to the graph bottom right-hand corner (FPR = 1 and TPR = 0). The AUC is always positive and ideally should be greater than 0.5. 

\section{PPV}
Also known as the class-specific accuracy scores of a classifier, both PPV and NPV account for the classifier precision regarding individual classes. The positive precision measures the fraction of correctly predicted positive samples over all samples marked to be positive, as we show in Equation~\ref{eq:ppv}~\citep{Zaki:2013}:

\begin{equation}\label{eq:ppv}
PPV = \frac{TP}{TP+FP} 
\end{equation}
\noindent where the TP are the true positives and the FP are the false positives. The PPV is measured on the scale \big[0:1\big], in which the higher the values, the better the classifier.

\section{NPV}
The negative precision, in turn, measures the fraction of correctly predicted negative samples over all samples marked to be negative, as we show in Equation~\ref{eq:npv}~\citep{Zaki:2013}: 

\begin{equation}\label{eq:npv}
NPV = \frac{TN}{TN+FN} 
\end{equation}
\noindent where the TN are true negatives and the FN are the false negatives. Similarly to the TPR, the NPV is also measured on the scale \big[0:1\big], where the higher the values, the better the classifier.

\section{ACC}
The overall accuracy, in turn, is a global score for classifier performance and also lies around the interval \big[0:1\big], in which the higher the score, the better the classifier. Accordingly, the overall accuracy accounts for the weighted mean of the class-specific accuracy scores, as we show in Equation~\ref{eq:acc}~\citep{Han:2006}.

\begin{equation}\label{eq:acc}
ACC = \frac{TP + TN}{TP + FN + FP + TN} 
\end{equation}
\noindent where the TP is the number of true positives, TN is the number of true negatives, FN are the false negatives, and FP are the false positives. It is worth mentioning that the ACC should not be used alone, since it can mask the real hit rates of individual classes in imbalanced data scenarios; however, we decided to keep it in our study for completeness purposes, since most authors use it.

\section{FAR}
Finally, the false alarm ratio represents the number of times a method forecasts events that are not observed. It is a complementary metric, and thus should be used with the positive recall to allow a better understanding of the observed false alarms. The FAR also lies around the interval \big[0:1\big]; however, the lower the score, the better the classifier. In Equation~\ref{eq:far} we show how to calculate the FAR~\citep{Jolliffe:2003}:

\begin{equation}\label{eq:far}
FAR = \frac{FP}{TP + FP} 
\end{equation}
\noindent where the FP is the number of false positives and the TP are the true positives. 

\begin{acks}
 This study was partly funded by the Coordenação de Aperfeiçoamento de Pessoal de Nível Superior (CAPES), Brazil -- Finance Code 001. Besides, we thank: (i) Federal Institute of Education, Science and Technology of Rio Grande do Sul (IFRS) -- Campus Feliz, for the cooperation with this research; (ii) NOAA/SWPC, for the provided data; (iii) Espa\c{c}o da Escrita, Pr\'o-Reitoria de Pesquisa, UNICAMP, for the language services provided; (iii) the reviewer, for the valuable comments and suggestions; 
\end{acks}
  
\bibliographystyle{spr-mp-sola}
\bibliography{library}  

\end{article} 

\end{document}